%% file: paper.tex
\documentclass[10pt]{article}
\pdfoutput=1

\usepackage{amsmath}
\usepackage{xcolor}

\usepackage{listings}
\usepackage{microtype}
\usepackage[T1]{fontenc}
\usepackage[inline]{enumitem}
\usepackage{tabularx}
\usepackage{float}
\usepackage{graphicx}
\usepackage{subfig}
\usepackage{booktabs}
\usepackage{todonotes}
\usepackage{siunitx}
\usepackage{xr-hyper}

\usepackage{multirow}
\usepackage{hyperref}

\usepackage[accepted]{icml2019}

\newcommand{\tool}[0]{Ithemal} 

\icmltitlerunning{\tool{}: Basic Block Throughput Prediction}

\definecolor{xxxcolor}{rgb}{0.8,0,0}
\definecolor{maroon}{RGB}{128,0,64}
\definecolor{darkgreen}{RGB}{0,128,64}
\newcommand{\IGNORE}[1]{}

\newcolumntype{L}[1]{>{\raggedright\arraybackslash}p{#1}}
\newcolumntype{C}[1]{>{\centering\arraybackslash}p{#1}}
\newcolumntype{R}[1]{>{\raggedleft\arraybackslash}p{#1}}

\usepackage{titlesec}

\titlespacing\section{0pt}{10pt plus 4pt minus 2pt}{0pt plus 2pt minus 2pt}
\titlespacing\subsection{0pt}{6pt plus 1pt minus 1pt}{0pt plus 1pt minus 1pt}
\titlespacing\subsubsection{0pt}{6pt plus 1pt minus 1pt}{0pt plus 1pt minus 1pt}

\begin{document}

\setlength{\abovecaptionskip}{-10pt}
\setlength{\belowcaptionskip}{-10pt}
\setlength{\intextsep}{7pt}
\setlength{\dbltextfloatsep}{8pt}
\setlength{\dblfloatsep}{8pt}

\setlength{\parskip}{0.2cm plus1mm minus1mm}

\setlength{\abovedisplayskip}{1pt}
\setlength{\belowdisplayskip}{3pt}

\twocolumn[
\icmltitle{\tool{}: Accurate, Portable and Fast Basic Block Throughput Estimation using Deep Neural Networks}

\begin{icmlauthorlist}
\icmlauthor{Charith Mendis}{MIT}
\icmlauthor{Alex Renda}{MIT}
\icmlauthor{Saman Amarasinghe}{MIT}
\icmlauthor{Michael Carbin}{MIT}
\end{icmlauthorlist}

\icmlaffiliation{MIT}{MIT CSAIL}

\icmlcorrespondingauthor{Charith Mendis}{charithm@mit.edu}

\icmlkeywords{Throughput Prediction, Deep Neural Networks}

\vskip 0.3in
]



\printAffiliationsAndNotice{}  

\input{abstract}
\input{introduction}
\input{motivation}
\input{model}
\input{dataset}
\input{evaluation}
\input{search}
\input{related}
\input{conclusion}
\input{acks}

\bibliography{references}

\bibliographystyle{icml2019}

\input{appendix}

\end{document}

%% file: abstract.tex
\begin{abstract}

Predicting the number of clock cycles a processor takes to execute a block of assembly instructions in steady state (the \emph{throughput}) is important for both compiler designers and performance engineers.
Building an analytical model to do so is especially complicated in modern x86-64 Complex Instruction Set Computer (CISC) machines with sophisticated processor microarchitectures in that it is tedious, error prone, and must be performed from scratch for each processor generation.
In this paper we present \tool{}, the first tool which \emph{learns} to predict the throughput of a set of instructions.
\tool{} uses a hierarchical LSTM--based approach to predict throughput based on the opcodes and operands of instructions in a basic block.
We show that \tool{} is more accurate than state-of-the-art hand-written tools currently used in compiler backends and static machine code analyzers.
In particular, our model has less than half the error of state-of-the-art analytical models (LLVM's llvm-mca and Intel's IACA).
\tool{} is also able to predict these throughput values just as fast as the aforementioned tools, and is easily ported across a variety of processor microarchitectures with minimal developer effort.

\end{abstract}

%% file: introduction.tex
\section{Introduction}

The {\em throughput} of a sequence of instructions---the number processor clock cycles taken to execute the sequence when looped in steady state---determines how fast those instructions can process data.
Accurately predicting the throughput of a basic block\footnote{In this work, we focus specifically on \emph{basic blocks}, sequences of instructions with no branches or jumps.} is an essential requirement in many systems, to be able to predict and optimize runtime performance.
For instance, constraint--based register allocation and instruction scheduling~\cite{comb-inst-scheduling} relies on accurate throughput estimations, as do learning--based techniques like genetic algorithm based register allocation~\cite{ga-register-allocation} and reinforcement learning based instruction scheduling~\cite{rl-inst-scheduling}.

The alternative -- measuring throughput on demand by executing the basic block -- is too expensive for most compilers and learning--based solutions.
In practice, most systems employ analytical models to predict throughput.
For instance, the LLVM compiler team~\cite{llvm} recently merged\footnote{ lists.llvm.org/pipermail/llvm-dev/2018-March/121490.html} a command-line tool, llvm-mca~\cite{llvm-mca}, that exposes a machine model for throughput estimation.
Intel has also released a closed-source machine code analyzer, IACA~\cite{iaca}, which relies on internal knowledge of Intel's processor design.
These models are typically an order of magnitude faster than measuring a basic block's throughput.
However, manually writing an accurate and complete model is tedious, error-prone, and exceedingly difficult without knowledge of the exact mechanisms of the processor.

In the hunt for {\em accuracy}, developers build complicated models which must make significant tradeoffs with the model's {\em portability} and {\em speed}.

\paragraph{\bf Accuracy.}
Modern x86-64 Complex Instruction Set Computer (CISC) processors contain many hardware optimizations that significantly complicate building accurate analytical models.
In order to implement an instruction set \emph{architecture} (ISA) like x86-64, processors actually implement an underlying \emph{micro}architecture, a physical implementation of the ISA specification.
Processors translate instructions from the ISA to instructions in the latent microarchitectural language (termed micro-ops), then execute those micro-ops.
The micro-ops may undergo optimizations such as
\textit{micro-op fusion}, in which micro-ops of different instructions may be combined together;
\textit{out-of-order execution}, in which instructions can be executed in any semantics--preserving order;
\textit{register renaming}, where false dependencies can be broken to enable more parallel execution;
and many more vendor-specific optimizations.
This makes the prediction problem highly complex and non-linear.

\paragraph{\bf Portability.}
While ISAs like x86-64 stay relatively stable, processor vendors release updated processor implementations with different microarchitectures every few years.
For example, Intel released the Haswell and Skylake microarchitectures in 2013 and 2015 respectively for the x86-64 instruction set.
Each microarchitecture of a processor family has its own quirks and intricacies.
Manually writing a throughput estimator to support different microarchitectures requires rewriting instruction tables, resource utilization charts, and modeling microarchitectural optimizations, all of which are tedious and error-prone.
This is complicated by the vast, incomplete, and incorrect documentation for many processors, where understanding of these behaviors has to be obtained by reverse-engineering the processor.
Ideally, the throughput estimator should be able to automatically capture such intricacies with minimal human intervention.

\paragraph{\bf Speed.} A throughput estimator also needs to be fast.
Compilers need to search through many code blocks before emitting the fastest version of a given instruction sequence.
Running the basic blocks to get the ground truth throughput requires sandboxing and many iterations of execution to arrive at a consistent steady-state throughput estimate, which is impractical for real-time systems.

\subsection{\tool{}: A Data Driven Approach}

In this paper we introduce {\it \tool{} (Instruction THroughput Estimator using MAchine Learning)}, which takes a novel data--driven approach to predicting throughput for a block of instructions, inspired by recent advances in Deep Neural Networks (DNNs).
\tool{} models the throughput estimation problem as a regression task and leverages a DNN to learn to predict throughput by using a large corpus of labeled data, mapping assembly sequences to real valued throughputs.
More concretely, \tool{} uses a a hierarchical multiscale RNN~\cite{hihi-hierarchical, chung-hierarchical, baraldi-hierarchical}, which generates an independent embedding for each instruction, then sequentially combines the instruction embeddings to predict throughput.

We show that \tool{}'s learned model is significantly more accurate than the analytical models, dropping the mean absolute percent error by more than 50\% across all benchmarks, while still delivering fast estimation speeds.

To generate high-quality predictions, \tool{} needs only training data and a specification of the ISA, including the specification of instructions and their explicit and implicit operands (for instance, the instruction \texttt{push rax} in x86-64 pushes the register \texttt{rax} on to the stack and also {\em implicitly} modifies the stack pointer register, \texttt{rsp}).
Unlike analytical models, \tool{} learns any salient microarchitectural details that contribute to throughput on its own, without any explicit specification or modeling.

In this paper, we present the following contributions:
\setlist{nolistsep}
\begin{itemize}[noitemsep]
\item{\bf Data--Driven Throughput Estimation.}
  We present \tool{}, the first system for data-driven basic block throughput estimation, using a hierarchical multiscale RNN.

\item{\bf Evaluation:}
  We demonstrate that \tool{} is more accurate than the state-of-the-art analytical throughput estimators, while still attaining fast estimation speeds.
  We also show that \tool{}'s design is portable across multiple processor microarchitectures.

\item{\bf DNN Architecture Exploration.}
  We demonstrate that the proposed hierarchical multiscale RNN outperforms many other choices, including other neural network architectures specifically tailored to mimic the dependency patterns of instructions within a basic block.

\item{\bf Open Source Implementation.}
We have open-sourced our implementation of \tool{} at \url{https://github.com/psg-mit/Ithemal} in the hope that performance engineers and compiler designers can use and improve upon our approach.

\end{itemize}

{\noindent}{\tool} demonstrates that future systems can leverage data--driven techniques to either augment or fully replace manually developed throughput estimators.

%% file: motivation.tex
\section{Motivating Examples}

Analytical modeling is difficult for many code sequences; consider examples (a)-(c), their actual throughput\footnote{Note that -- following convention -- we define throughput to be the number of clock cycles taken to execute a basic block; this is actually the reciprocal of the standard definition of throughput. We also report throughput for 100 iterations of a given basic block.}, and their associated throughput predictions in Table~\ref{tbl:motivation}.
These flawed predictions occur in spite of many hours spent engineering detailed models of underlying microarchitectural details.
In contrast, \tool{}'s data driven approach intrinsically learns accurate predictions from the ground truth data.

\begin{table}[h!]
  \centering
  \includegraphics[width=\columnwidth]{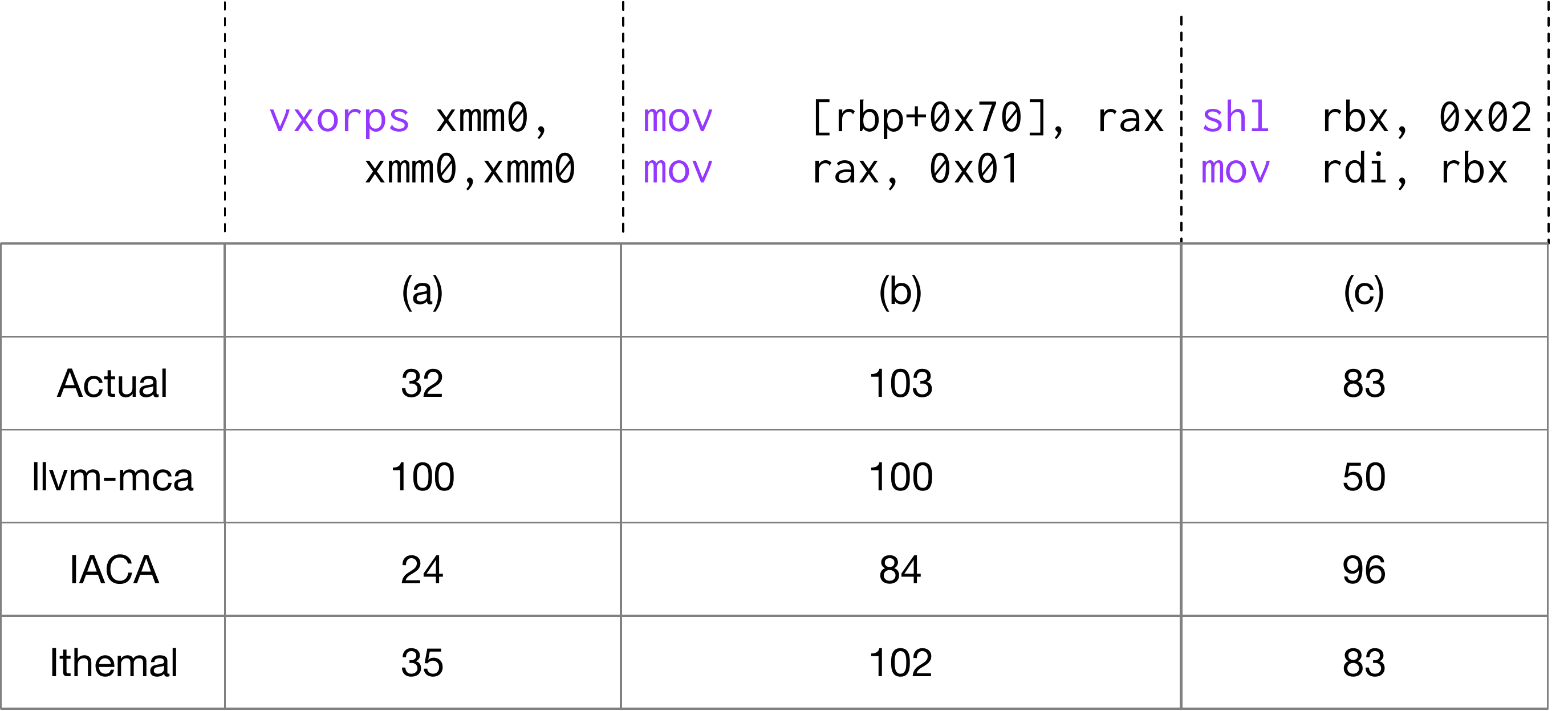}
  \vspace{-20pt}
  \caption{Example x86-64 assembly code sequences (Intel syntax) and associated throughput predictions, in clock cycles}
  \label{tbl:motivation}
\end{table}

\begin{figure*}[ht!]
  \centering
  \includegraphics[width=\textwidth]{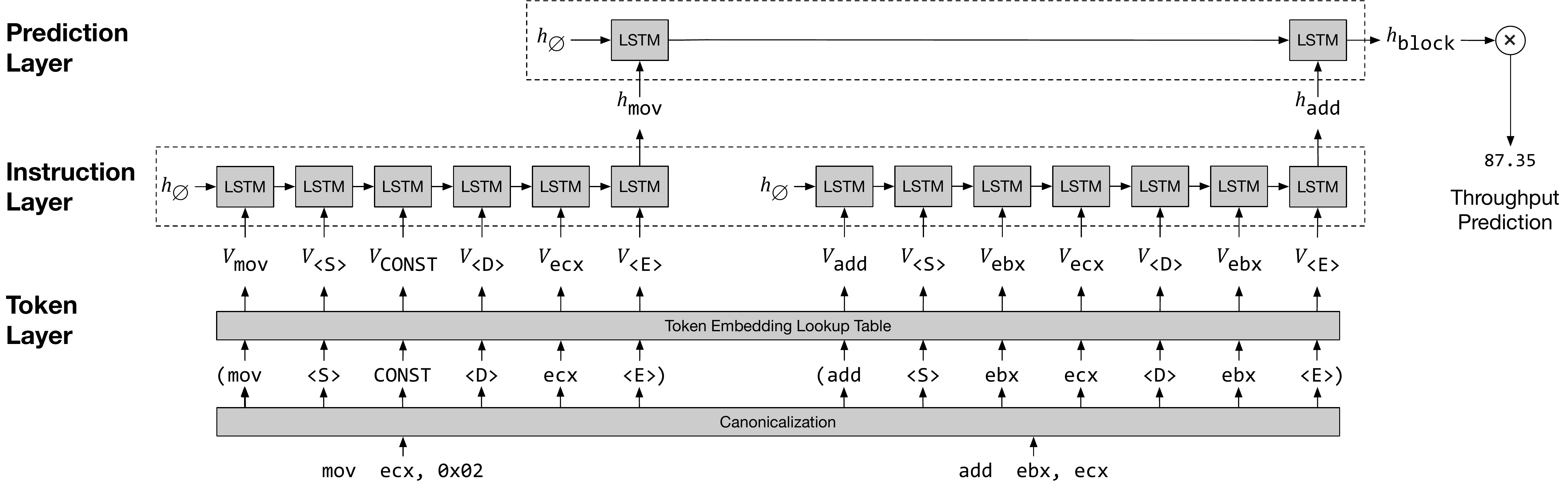}
  \vspace{-20pt}
  \caption{\tool{} System Architecture}
  \label{fig:overall}
\end{figure*}

\paragraph{Implementation Errors:}
Intel provides extensive documentation of its microarchitectural implementation that enables developers to build performance models for assembly code.
However, the sheer volume of implementation details makes it challenging to deliver a complete model.

Sequence (a) shows a single instruction sequence that zeros out the vector register \texttt{xmm0}.
Zeroing out registers is so common that Intel processors execute these instructions using a faster, optimized data path -- separate from the normal instruction execution path.
IACA closely predicts the measured value but llvm-mca's predictions are much farther off, because it does not model this optimization.

Sequence (b) shows a pair of \texttt{mov} instructions with a measured throughput of $103$.
IACA and LLVM both find an execution schedule which would predict a throughput of $100$ cycles; however, IACA also identifies a micro-op fusion opportunity, and therefore predicts $84$ cycles.
This optimization opportunity does not manifest in the observed timing numbers.

\tool{}, which is driven by actual performance data, learns to closely predict both of these values without any explicit error-prone encoding of Intel's optimizations.

\paragraph{Vendor Documentation Errors:}
The sheer volume of implementation details also means that Intel's documentation can be incorrect.
Tools that faithfully adhere to the documentation can therefore still be incorrect.
Sequence (c) is a short sequence with a data dependency that is bypassed within the processor pipeline: the \texttt{mov} instruction does not consume many additional clock cycles over that of \texttt{shl}.
The throughput of this basic block is therefore dominated by the throughput of \texttt{shl}.
However, the throughput value that Intel provides in its documentation (50 cycles) assumes that there are no dependencies.
Therefore, while IACA -- Intel's own tool -- closely predicts the value, llvm-mca is incorrect because it uses the dependency-free throughput value.
In comparison, \tool{} closely predicts the actual throughput because it works with actual performance data.

%% file: model.tex
\section{Model Architecture}
\label{sec:model}

Figure~\ref{fig:overall} presents the high-level design of \tool{}'s approach.
We model the problem of throughput estimation as a regression problem: given the assembly input, \tool{} predicts the throughput of the instruction sequence as a real-valued number.
At the core of \tool{} is a hierarchical multiscale RNN~\cite{dag-rnn1, dag-rnn2} that sequentially processes all instructions in the basic block and outputs an embedding, which \tool{} then uses to directly estimate the throughput.
Altogether, we decompose the end-to-end model into the following stages: \emph{canonicalization}, \emph{embedding} and \emph{estimation}.

\subsection{Canonicalization}
\label{sec:canonicalization}

The canonicalization stage converts the assembly input into a more structured form, dictated by the syntax of the assembly instructions.
\tool{} takes a compiled assembly block, disassembles it, and maps it to a list of instructions.
Each instruction consists of a list of tokens representing its operation code (opcode, e.g. \texttt{add}), source operands, and destination operands, separated by distinguished delimiter tokens.

For example, consider the instruction \texttt{mul ecx}, which multiplies the value in register \texttt{ecx} with \texttt{eax}, and places the result into registers \texttt{edx} and \texttt{eax}.
Note that the source operand \texttt{eax} and both of the destination operands \texttt{eax} and \texttt{edx} are implicit in the Intel syntax \texttt{mul ecx}.
The final canonicalized set of tokens for the instruction is:
$$(\texttt{mul},\,\textit{<S>},\,\texttt{eax},\,\texttt{ecx},\,\textit{<D>},\,\texttt{edx},\,\texttt{eax},\,\textit{<E>})$$
where the bracketed tokens are the delimiters representing the break between the opcode, source, and destination operands.

Assembly code permits more than just register operands, such as constants and memory operands.
We map all constants (e.g. integer constants, memory addresses, etc.) to a single \texttt{CONST} token.
We demarcate memory operands (consisting of a base address, and an optional offset and displacement) by surrounding them with \texttt{<M>} and \texttt{</M>} delimiter tokens.
We present the full canonicalization scheme in Appendix~\ref{app:token-grammar}.

\subsection{Embedding}
\label{sec:nnembeddings}

{\tool}'s embedding stage takes a canonicalized token stream of instructions, and for each instruction produces an \emph{embedding}: a representation of an instruction as a real-valued vector in a high-dimensional space.
The first step is the \textit{token layer}, which maps a given token to an embedding.
We implement the token layer by mapping each token in the sequence to an $n$-dimensional vector by learning a linear transformation of the one-hot token vectors (this is equivalent to learning a lookup table).

\tool{} then maps the sequence of token embeddings to an embedding for each instruction in the basic block.
We call this the \textit{instruction layer}.
Because each instruction can have a variable number of tokens depending on its number of source and destination operands, the size of the input to the embedding stage is variable.
We therefore implement the instruction layer with a sequential Recurrent Neural Network (RNN) architecture with Long Short Term Memory (LSTM)~\cite{lstm} cells.

Figure~\ref{fig:overall} presents the operation of our RNN-based instruction embedding approach on a small example.
The bottommost row shows the original assembly input.
The second row shows the sequence of tokens for each instruction.
The third row (the token layer) shows the sequence of {\em token embeddings}, e.g. $v_\texttt{mov}$, which are mapped directly from each syntactic token.
The fourth row (the instruction layer) shows the application of an LSTM to reduce the token embedding sequence into the final instruction embedding, $h_{\texttt{mov}}$.

\subsection{Prediction}
\label{sec:nnoverview}

The final prediction comes from the \textit{prediction layer}, which maps a basic block (a sequence of instruction embeddings) to a throughput value.
This is again implemented with an RNN with LSTM cells, which has entirely disjoint weights from the LSTM in the instruction layer.
This corresponds to the topmost layer in Figure~\ref{fig:overall}.
Using the final output from the instruction LSTM ($h_{\texttt{block}}$), \tool{} predicts the basic block's throughput with a linear layer.
Specifically, {\tool} computes $w \cdot h_{\texttt{block}} + b$, where $w$ is a learned weight vector and $b$ is a bias.
This produces a final real-valued number that represents the network's throughput prediction.

The hierarchical combination of the RNN in the instruction layer and the RNN in the prediction layer has several benefits over a non-hierarchical model:
\begin{itemize}
\item Memory and backpropagation paths are significantly shorter than a model using a non-hierarchical (i.e., single) RNN.
  The average length of a block in our dataset is $6.04$ instructions, and the average length of an instruction is $7.97$ tokens.
  The average length of a token-level RNN across the entire basic block would instead be about $48$ RNN cell applications, more than three times as long as a path through the hierarchical RNN.
\item
  Instructions are embedded atomically: the prediction layer is only able to generate throughput estimates at specific points in the overall token stream, i.e. after complete instructions.
  This means that the network does not have an obligation to produce states that correspond to predictions at points in between instructions.
\end{itemize}
We compare this hierarchical architecture against other architecture choices in Section~\ref{sec:exploration}, showing the efficacy of \tool{}'s hierarchical architecture.

%% file: dataset.tex
\section{Data and Training}

\begin{table*}[ht!]
  \small
  \begin{center}
\begin{tabular}{ |l|p{8cm}|C{1.5cm}|C{1cm}| }
  \hline
  \textbf{Benchmark suite} & \textbf{Description} & \textbf{\#Total Blocks} & \textbf{\#Unique Blocks}  \\
  \hline
  Linux Shared Libraries & linux loaders, standard library and other utilities & 313846 & 103977\\
  \hline
  \multirow{3}{*}{SPEC2006~\cite{SPEC06}}  & benchmark suite with compilers, chess engines, video compression and various simulation applications. Commonly used for benchmarking compilers & \multirow{3}{*}{247047} & \multirow{3}{*}{141051} \\
  \hline
  SPEC2017~\cite{SPEC2017} & similar to SPEC2006, but with a larger variety  & 616899  & 234588\\
  \hline
  NAS~\cite{NAS} & benchmarks with stencil computations (dense loops) & 3935 & 1813\\
  \hline
  polybench-3.1~\cite{polybench} & polyhedral compilation test suite (dense loops) & 1900 & 859\\
  \hline
  TSVC~\cite{tsvc} & suite for testing compiler auto-vectorization & 5129 & 2350\\
  \hline
  cortexsuite~\cite{cortexsuite} & computer vision workloads including neural networks & 6582 & 3968\\
  \hline
  \multirow{2}{*}{simd~\cite{simd}} & heavily hand vectorized image processing library (exposes lot of SSE2, AVX, AVX2 variants)  & \multirow{2}{*}{212544} & \multirow{2}{*}{25462}\\
  \hline
  \multirow{2}{*}{compilers/interpreters} & clang~\cite{llvm} and different versions of python (2.7,3.5) & \multirow{2}{*}{2746275}  & \multirow{2}{*}{924663}\\
  \hline
  end user applications & gimp filters, firefox, open-office, rhythmbox, etc. & 83555 & 35513 \\
  \hline
  \hline
  Full Dataset & & 4237712 & 1416473 \\

  \hline
\end{tabular}
  \end{center}
  \vspace{-10pt}
  \caption{Composition of the dataset for Haswell, showing the total number of basic blocks per benchmark as well as the unique number of blocks after de-duplicating repeated blocks on a per-benchmark basis. Note that the \#Unique Blocks on the full dataset is not equal to the sum of unique blocks per individual benchmark, as we de-duplicate across all benchmarks.}
  \label{tbl:bench}
\end{table*}

We collected a dataset of basic blocks from well-known programs and benchmark suites, and timed them with a procedure that matches the assumptions of the baseline analytical models.
We then train \tool{} using standard supervised learning techniques.
\subsection{Dataset}
\label{sec:dataset}

Table~\ref{tbl:bench} summarizes the set of applications in our dataset.
We designed the dataset to include a diverse set of applications with different performance characteristics
while covering a wide range of x86-64 instructions.
It consists of performance critical applications used for benchmarking compiler optimizations
as well as end user applications used in day-to-day computing.

To extract each application's basic blocks, we first compile each application using GCC 4.9.4 with the \texttt{-O3} optimization level targeting an Intel Haswell processor. 
Next, we use Dynamorio~\cite{dynamorio}, a dynamic binary instrumentation tool, to dump the encoded bytes of the executed x86-64 basic blocks.
We execute the benchmarks using the standard inputs provided by the benchmark suites.
Next, we de-duplicate the dataset by removing basic blocks with same encoded byte patterns. This step is important to eliminate repeated occurrences of basic blocks created by code shared through common header files and by common compilation patterns.

\subsection{Throughput Profiling}

IACA and llvm-mca predict the steady-state throughput of a basic block, under the assumptions that all memory accesses result in L1 cache hits and that the execution environment is non-preemptive.
To collect compatible throughput numbers, we profile the execution of a loop that executes each basic block in isolation 100 times (enough to reach the steady-state behavior; 100 iterations is also the default value used by llvm-mca).
We measure throughput in terms of clock cycles using a script that we have developed that is similar to Agner Fog's timing script\footnote{https://www.agner.org/optimize/testp.zip}. Agner Fog's timing script is commonly used for validating individual instruction throughputs.
Our timing script additionally ensures that almost all memory accesses have a L1 cache hit.
Additionally, we measure L1 instruction and data cache misses and software context switches to detect and filter out invalid executions that do not conform to the assumptions made by IACA and llvm-mca in their predictions.

Using this methodology, we collected valid throughput values for the Intel Ivy Bridge (Intel(R) Xeon(R) CPU E5-2695 v2), Haswell (Intel(R) Xeon(R) CPU E5-2680 v3) and Skylake (Intel(R) Xeon(R) W-2123 CPU) microarchitectures.
Data collection takes approximately 4-5 days for each microarchitecture.
Table~\ref{tbl:bench} shows the breakdown of basic block counts for each benchmark for Haswell microarchitecture in total as well as after de-duplicating repeated basic blocks on a per benchmark basis.
The final Haswell dataset, which is de-duplicated across benchmarks, constitutes 1,416,473 unique basic blocks.

\subsection{Training and Methodology}
\label{sec:training}

We implemented our neural network model in PyTorch (0.4.0a0+59bda9a).
The learnable parameters in \tool{} include the token embeddings, the token LSTM and instruction LSTM parameters, and the affine coefficients in the final linear layer.
For our loss function we use a normalized error metric, based on the L1 norm:
$$\mathcal{L}(\texttt{pred}, \texttt{actual}) = \frac{|\texttt{pred} - \texttt{actual}|}{\texttt{actual}}$$
We randomly assign 80\% of the collected blocks to the train set and 20\% to the test set.
We use Asynchronous Stochastic Gradient Descent~\cite{SGD,hogwild} to train the model.
Our full training regime is detailed in Appendix~\ref{app:hyperparams}.

%% file: evaluation.tex
\section{Evaluation}
\label{sec:evaluation}

\begin{figure*}[t!]
	\centering
	\subfloat[][\tool{}]{
		\includegraphics[width=0.33\textwidth]{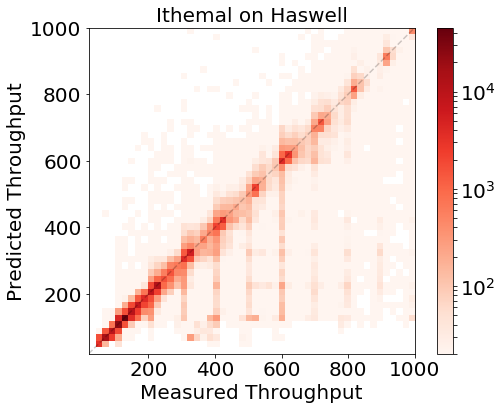}
	}
	\subfloat[][llvm-mca]{
		\includegraphics[width=0.33\textwidth]{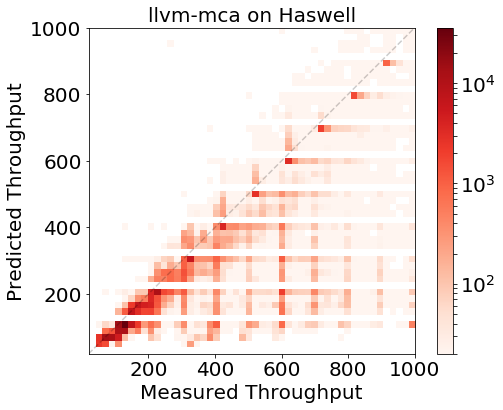}
	}
	\subfloat[][IACA]{
		\includegraphics[width=0.33\textwidth]{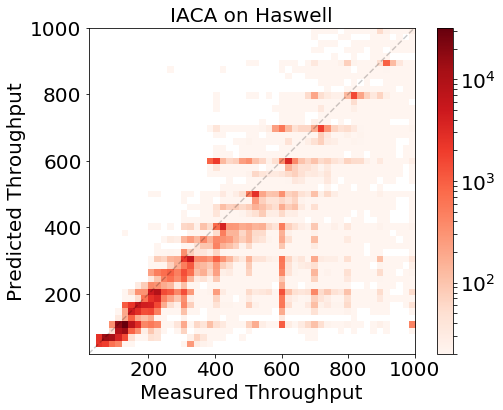}
	}

	\caption{Heatmaps for measured and predicted throughput values under different models for basic blocks with measured throughput values less than 1000 cycles (Haswell)}
	\label{fig:accuracy}
\end{figure*}

We have evaluated \tool{} against two state-of-the-art, hand-written analytical models: IACA~\cite{iaca} (v3.0-28-g1ba2cbb) and llvm-mca~\cite{llvm-mca} (LLVM 8.0.0).
Both of these models are designed to model the complexities of modern processors (including pipelining, superscalar, and out-of-order units).
We show that our data--driven model beats the accuracy of these sophisticated hand-written models (Section~\ref{sec:accuracy}) while maintaining just as fast prediction speeds (Section~\ref{sec:speed}).
Further, we show that our approach is portable across different microarchitectures in Section~\ref{sec:portability} by showing that \tool{} learns a model that outperforms IACA and llvm-mca without any neural network architecture or hyperparameter modifications.

\subsection{Accuracy}
\label{sec:accuracy}

We evaluate the accuracy of each model against the actual throughput values for Intel's Haswell, Ivy Bridge, and Skylake microarchitectures.
The version of IACA we use does not support throughput estimation for Ivy Bridge; we therefore evaluate accuracy only for \tool{} and llvm-mca for Ivy Bridge.
We prepared datasets for each microarchitecture according to the methodology described in Section~\ref{sec:training}.

Table~\ref{tbl:accuracy} presents the results of our accuracy comparison.
We report the average error with respect to the ground truth of each tool for each microarchitecture.
We also report both the Spearman and Pearson correlation of each tool's predictions with the ground truth.

\tool{} is more accurate in its throughput predictions for basic blocks across all three microarchitectures.
Our model's predictions are closer to the ground truth than both IACA and LLVM in 74\% of the blocks in the Haswell test set.
\tool{}'s predictions also have a higher correlation with the ground truth values for both the Spearman (rank correlation) and Pearson (linear correlation) metrics.
The higher Spearman correlation is especially useful because it directly corresponds to higher utility for use within an optimizing compiler (such as an instruction scheduling pass).
Specifically, compilers typically only need to determine which of several configurations of a basic block is the fastest, and do not calculate each block's absolute performance.

Figure~\ref{fig:accuracy} presents three heatmaps relating actual and predicted values in for basic blocks with throughputs less than 1000 cycles for each prediction method (representing $95\%$ of our dataset).

\begin{figure}[h!]
  \centering
  \includegraphics[width=\columnwidth]{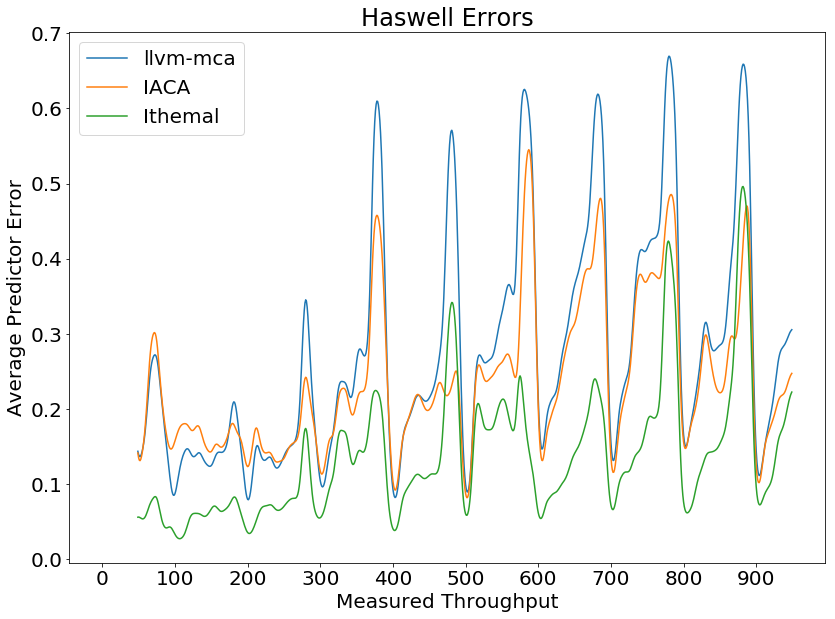}
  \includegraphics[width=\columnwidth]{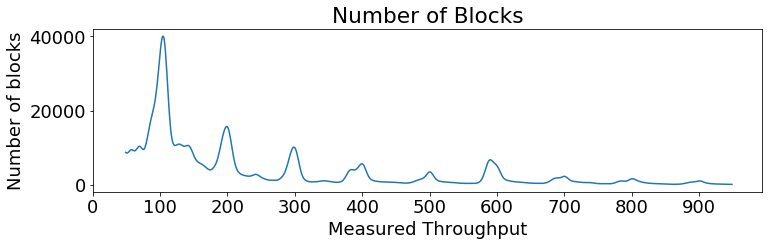}
    \caption{Average error across throughputs for different estimation methods for basic blocks with throughput values less than 1000 cycles (Haswell)}
    \label{fig:prederror}
  \end{figure}

To generate each heatmap, we binned the actual and predicted data into axis-aligned bins of width and height 20 cycles.
The color in each bin represents the count of blocks in that bin.
A perfect estimator would have all points along the line $y=x$ (shown as a faint grey, dashed line on the heatmaps), since the predicted throughputs would always match the measured throughputs.
We see a higher density near the identity line for \tool{}, compared to both llvm-mca and IACA.
Both llvm-mca and IACA also have more horizontal banding, representing more predictions of the same throughput value for different blocks that do actually have different behaviors.
Figure~\ref{fig:prederror} shows the average error of each system across a range of throughputs.
Compared to llvm-mca and IACA, \tool{} is better for blocks of almost all sizes.
All estimators struggle with blocks with measured throughputs just below peaks
in the measured throughput distribution. We hypothesize that these blocks correspond to special
microarchitectural optimizations that llvm-mca and IACA do not model. Ithemal also struggles
some with these rare blocks, but still outperforms both analytical estimators.

\begin{table}[h!]
  \small
  \begin{center}
\begin{tabularx}{0.48\textwidth}{ |p{1.4cm}|l|l|C{1.5cm}|C{1.2cm}| }
  \hline
  Micro\-architecture & Method & Error & Spearman Correlation & Pearson Corr. \\
  \hline
  Ivy Bridge  & llvm-mca & 0.181 & 0.902 & 0.777 \\
              & \tool{}  & \textbf{0.089} & \textbf{0.955} & \textbf{0.913} \\
  \hline
  Haswell  & llvm-mca & 0.200 & 0.890 & 0.790 \\
           & IACA  & 0.209 & 0.917 & 0.833 \\%
           & \tool{}  & \textbf{0.089} & \textbf{0.960} & \textbf{0.918} \\
  \hline
  Skylake  & llvm-mca & 0.239 & 0.852 & 0.729 \\
           & IACA  &  0.167 & 0.926 & 0.835 \\
           & \tool{}  & \textbf{0.079} & \textbf{0.960} & \textbf{0.895} \\
  \hline
\end{tabularx}
  \end{center}
  \vspace{-10pt}
  \caption{Average error for different models and microarchitectures}
  \label{tbl:accuracy}
\end{table}

\subsection{Speed}
\label{sec:speed}

Table~\ref{tbl:speed} presents the results of our evaluation of \emph{estimator throughput}: the number of instructions able to be timed per second for each estimator.
We calculate this by measuring the number of basic blocks each tool can time per second, and multiplying that by the average number of instructions per basic block.
In the last row, we also show the corresponding estimator throughput if we instead measure the ground-truth throughput for a given block.
We measured these estimator throughputs on the Haswell test set on a machine with an Intel Xeon E5-2680 CPU.

\tool{} is as fast as llvm-mca and IACA in our measurements, and is significantly faster than empirical evaluation of basic blocks.
It is worth noting that llvm-mca and IACA can both also output diagnostic information about basic blocks, and also that empirical evaluation of ground-truth data could be sped up by running fewer repeated measurements (it may be possible that as few as 2 or 3 measurements would suffice in some contexts).
However even with these qualifications, we show that \tool{} functions as an equivalently performant and more accurate drop-in replacement for llvm-mca and IACA in systems which only need throughput estimations, while still performing significantly faster than empirical evaluation.

\begin{table}[h!]
  \small
  \begin{center}
\begin{tabular}{ |l|c| }
  \hline
  Method & Throughput (Instructions / second)  \\
  \hline
   llvm-mca & 492  \\  
   IACA & 541  \\ 
   \tool{} & 560 \\ 
   Empirical execution & 13 \\ 
  \hline
\end{tabular}
  \end{center}
  \vspace{-10pt}

  \caption{Estimation throughputs for different estimators measured in instructions per second}
  \label{tbl:speed}
\end{table}

\subsection{Portability}
\label{sec:portability}

We designed and trained \tool{} on Haswell and validated our architecture and hyperparameters by re-training on Skylake.
Without any changes to its structure or training regime, we then trained and evaluated \tool{} on the Ivy Bridge dataset.
Table~\ref{tbl:accuracy} summarizes the average errors for each microarchitecture.
\tool{} learns to estimate throughput values for each microarchitecture with a maximum average error of 0.089 across all datasets.
The hand-written models exhibit a minimum average error of 0.167.

In sum, \tool{} provides state-of-the-art prediction performance; its results beat the baselines across the board.
Moreover, \tool{} does so without requiring a user to provide information about the processor's underlying microarchitecture, whereas these analytical models require significant re-engineering for each microarchitecture of interest.

%% file: search.tex
\section{Neural Network Architecture Exploration}
\label{sec:exploration}

We evaluated a number of neural network architectures with varying levels of structure and complexity before arriving at {\tool}'s network architecture (Section~\ref{sec:model}).

\begin{figure}
    \centering
    \includegraphics[width=\linewidth]{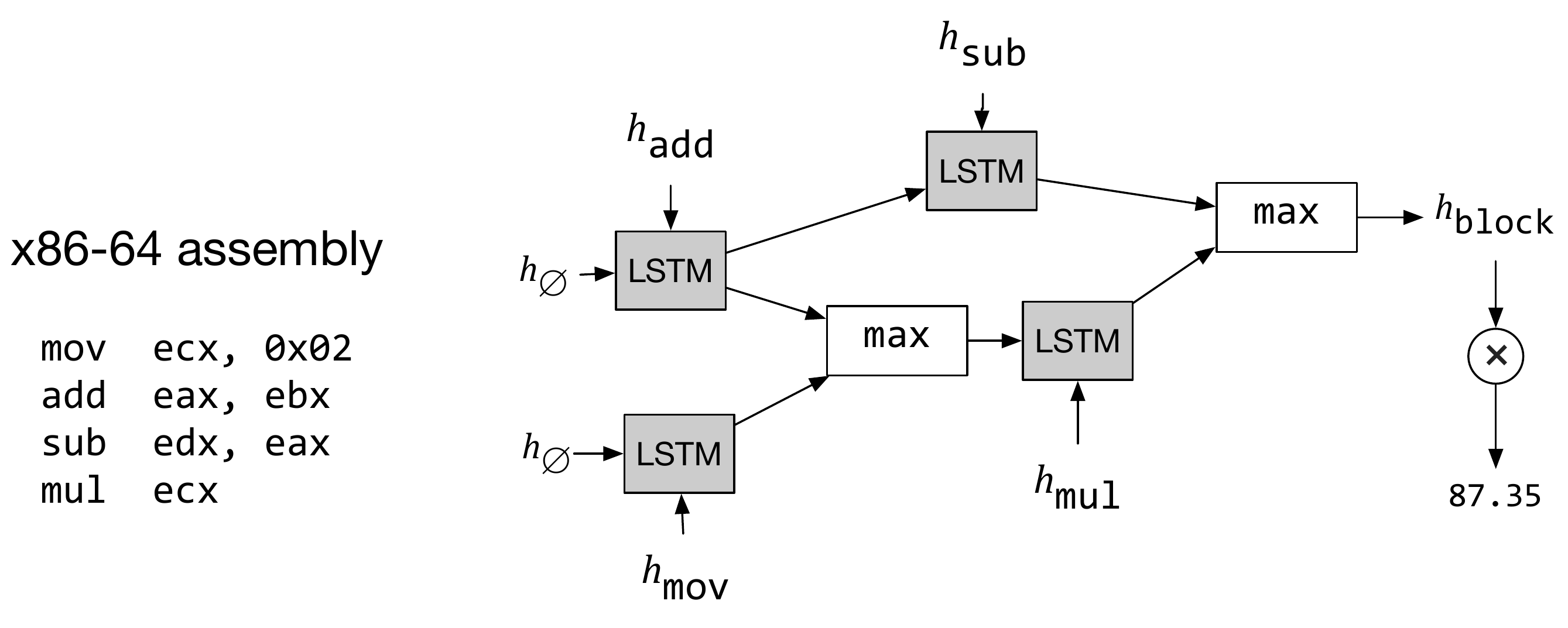}
    \vspace{-20pt}
    \caption{The DAG-RNN Architecture}
    \label{fig:diff-networks}
\end{figure}

Figure~\ref{fig:diff-networks} shows a DAG-RNN~\cite{dag-rnn1, dag-rnn2}.
Instructions are embedded identically as to in \tool{} (i.e. the token layer and instruction layer remain the same).
However rather than running an RNN sequentially over the instructions in the prediction layer, the DAG-RNN constructs a dependence graph of the instructions in the basic block, with a directed edge between a pair of instructions if the second instruction depends on an output of the first instruction.
Because an instruction can depend on multiple previous instructions, we apply an element-wise max to reduce the states feeding in to a given instruction.
Then, the DAG-RNN applies an LSTM cell, using the generated instruction embedding as the input, and the result of the element-wise max as the input state.
To generate the final prediction, we take an element-wise max of the output states of all leaf instructions (all instructions with no dependents), and pass the result through a linear layer.

The DAG-RNN is inspired by the theoretical behavior of a perfect out-of-order processor: the throughput of a basic block running on a perfect out-of-order processor is equivalent to the throughput of the longest path that must be serially executed in that basic block.
Using a DAG-RNN implicitly encodes this prior, by only allowing information to propagate through the paths that must be serially executed in the block.

We also tested a simple token-level RNN with LSTM cells, which has a similar base architecture as \tool{}, but without the topmost prediction layer.
Instead, this model sequentially consumes all tokens in a basic block making no explicit distinction between instructions, giving a baseline measure for the efficacy of \tool{}'s hierarchical model.
The full architecture diagram for the token-level RNN is shown in Appendix~\ref{app:token-rnn-architecture}.

\begin{figure}[h!]
  \includegraphics[width=\linewidth]{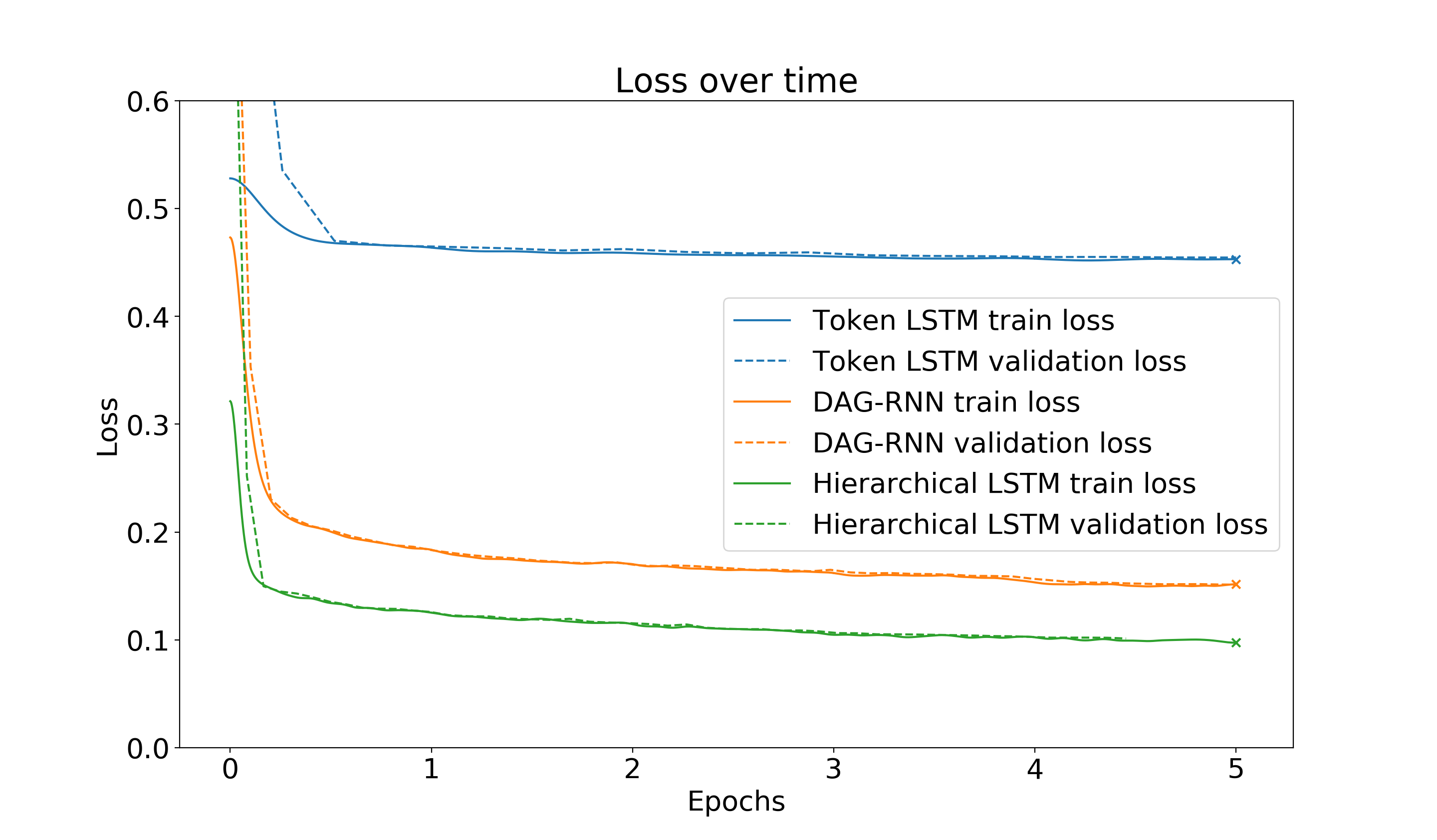}
  \vspace{-20pt}
  \caption{Learning curves for the models}
  \label{fig:learn}
\end{figure}

\paragraph{\bf Results.}
Figure~\ref{fig:learn} shows the training and validation loss for each model across the first five epochs.
It is clear that the hierarchical LSTM is the best model among these three.
The sequential LSTM performs the worst by far, motivating the need to process individual tokens and instructions at multiple scales.
The fact that the DAG-RNN performs worse than the hierarchical LSTM implies that the exact ordering of instructions in a basic block does matter, not just the dependency chains.
This aligns with the fact that instruction scheduling optimizations in compilers do result in changed performance, despite the underlying dependency graph being the same.
While the perfect out-of-order execution model is a reasonable approximation, modern processors do in fact have some serial behavior, which a sequential model is able to capture.

%% file: related.tex
\section{Related Work}

\paragraph{Hierarchical Multiscale RNNs}
Hierarchical RNNs are a classic technique in sequence prediction domains~\cite{hihi-hierarchical}.
\tool{} uses a hierarchical multiscale RNN as the core of its prediction methodology, similar to many other proposed hierarchical RNN models.
\tool{} has some inherent advantages over other models from the literature: as opposed to the methodology proposed in~\cite{chung-hierarchical}, the hierarchical structure in instruction embedding and throughput prediction is explicit rather than latent.
\tool{}'s structure is instead most similar to boundary-aware hierarchical RNNs, such as the one presented in~\cite{baraldi-hierarchical} for video captioning based on a hierarchy of frames and scenes.

\paragraph{DAG-RNNs and Graph Neural Networks}
Neural networks with generic graph based structures have been used in NLP tasks to model relations among words in sentences ~\cite{acl:rel-extraction, arxiv:rel-extraction}.
Programs also can be represented using Gated Graph Neural Networks~\cite{ggnn}, to perform high-level tasks such as variable naming, or identifying variable misuses.
\citet{binary-similarity}~uses graph neural networks to find binary similarity between different execution platforms.
In experimenting with a DAG-RNN~\cite{dag-rnn1,dag-rnn2}, we hoped to be able to take advantage of the program semantics of a basic block, although that approach does not perform as well in our collected datasets.

\paragraph{Analytical Models for Throughput and Runtime Estimation}

Apart from state-of-the-art tools like llvm-mca and IACA, other analytical models exist for throughput estimation~\cite{taha-tp} of instructions.
OSACA~\cite{laukemann-osaca} is an open source analytical model similar to llvm-mca and IACA, which automates some of the collection of the tabular data which is plugged into the model.
There are also analytical models such as~\cite{chen-tp-multi} to estimate throughput for multithreaded programs.
Cycle--accurate simulators such as ZSim~\cite{sanchez-sim} and Marss~\cite{marss-sim} have a high start-up cost and are more suited for coarse grained simulations.

Coarser analytical models exist for predicting program runtimes~\cite{park-prediction}.
Work has also been done in developing analytical models to predict performance of restricted classes of programs.
For example, work on predicting parallel program runtimes include~\cite{rugina-parallel, adve-parallel, hartleb-parallel, blanco-parallel, fahringer-parallel}
and work on predicting worst case execution times include~\cite{ferdinand-wcet,li-wcet}.

All of these models require detailed processor modeling and considerable human development effort.

\paragraph{Learned Models for Throughput and Runtime Estimation}

There has been work on developing machine learning--based models for absolute and relative runtime estimation.
\cite{sparse-reg}~introduces sparse polynomial regression to predict execution time of programs by using a set of hand--crafted features of high level programs.
\citet{suif-speedup}~uses neural networks with hand-crafted features to estimate the speedup between two code sequences.
GameTime~\cite{gametime, gametime2} uses SMT solvers to generate inputs and game theoretic approaches to predict the distribution of runtimes of programs.

These models require manual feature engineering, and runtime predictions are done at a coarser granularity (e.g. at the full program level).
In contrast, \tool{} automatically learns how to predict throughput of basic blocks with minimal architectural knowledge embedded into the model.

\paragraph{Microarchitectural Predictions}

Similar to basic block throughput estimation, various microarchitectural prediction tasks have been explored with machine learning.
For example, RNN models can be used for predicting memory access~\cite{learning-memory}, and perceptron models can be used for branch prediction~\cite{perceptron-branch}

%% file: conclusion.tex
\section{Conclusion}

We present {\tool}, a data--driven system for basic block throughput estimation.
{\tool}'s accuracy surpasses that of state-of-the-art, hand-written analytical models; it achieves its accuracy by leveraging a deep neural network designed to capture the behavior of modern processors.
{\tool} demonstrates that future compilation and performance engineering tools can be augmented with data--driven approaches to improve their performance and portability, while minimizing developer effort.

%% file: acks.tex
\section*{Acknowledgments}

We would like to thank Yishen Chen, Ajay Brahmakshatriya for their help in creating the datasets, and all reviewers for insightful comments and suggestions.
We would also like to thank Ond\v{r}ej S\'{y}kora and Jon Orwant for their many helpful discussions.
This research was supported by DARPA D3M Award \#FA8750-17-2-0126, NSF Grant CCF-1751011, DARPA/Nvidia Symphony Award \#HR0011-18-3-0007 and DARPA Award \#HR001118C0059.
Any opinions, findings, and conclusions or recommendations expressed in this material are those of the authors and do not necessarily reflect the views of the funding agencies.

%% file: appendix.tex
\appendix

\section{Canonicalization Scheme}
\label{app:token-grammar}

We demarcate memory operands (consisting of a base address, and an optional offset and displacement) by surrounding them with \texttt{<M>} and \texttt{</M>} delimiter tokens.

The following is the full grammar for the token strings, as described in Section~\ref{sec:canonicalization}.
\begin{align*}
  \textit{<block>} &::= \textit{<instr>}+\\
  \textit{<instr>} &::= \textit{opcode}\ \texttt{<S>}\ \textit{<opnd>}\hspace{-.2em}*\ \texttt{<D>}\ \textit{<opnd>}\hspace{-.2em}*\ \texttt{<E>} \\
  \textit{<opnd>} &::= \textit{register} \; | \; \texttt{<M>}\ \textit{register}\hspace{-.2em}+ \texttt{</M>} \; | \; \texttt{CONST}
\end{align*}

\section{Training Hyperparameters}
\label{app:hyperparams}

Here we present the hyperparameters for the model as described in Section~\ref{sec:model}.
All vectors, including the embedding width, hidden, and output states have width 256.
We train our models using asynchronous SGD, with a batch size of $4$, and $6$ parallel trainers.
The initial learning rate is $0.1$, and after the first $2$ epochs, it decreases by a geometric factor of $1.2$ every epoch.
We use the default PyTorch formulation for momentum (i.e. not Nesterov momentum) with $\beta=0.9$.
Each parallel trainer samples without replacement from the dataset until all training data is exhausted.
If a trainer hits a \texttt{NaN} gradient, that trainer is halted for the remainder of the epoch, and the elements in that trainer's batch are dropped for the epoch.
At the beginning of the next epoch, all trainers are restarted.
Training halts once all trainers are halted and an epoch cannot be completed.

\section{Heatmaps of Different Prediction Methods}
\label{app:heatmaps}

\begin{figure*}
	\centering
	\subfloat[][\tool{}]{
		\includegraphics[width=0.33\textwidth]{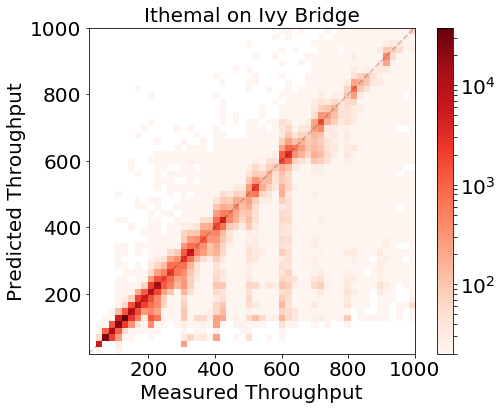}
	}
	\subfloat[][llvm-mca]{
		\includegraphics[width=0.33\textwidth]{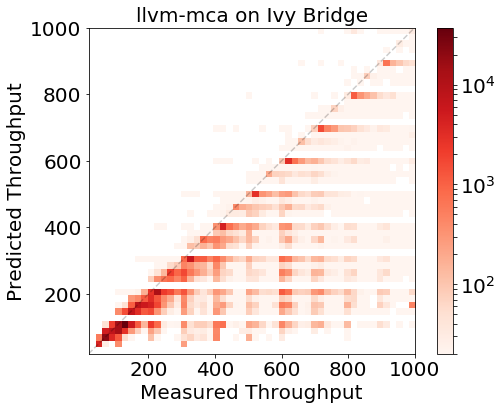}
	}\\
	\subfloat[][\tool{}]{
		\includegraphics[width=0.33\textwidth]{figures/heatmap_ithemal_haswell.png}
	}
        \subfloat[][llvm-mca]{
		\includegraphics[width=0.33\textwidth]{figures/heatmap_iaca_haswell.png}
	}
        \subfloat[][IACA]{
		\includegraphics[width=0.33\textwidth]{figures/heatmap_iaca_haswell.png}
	}\\

	\subfloat[][\tool{}]{
		\includegraphics[width=0.33\textwidth]{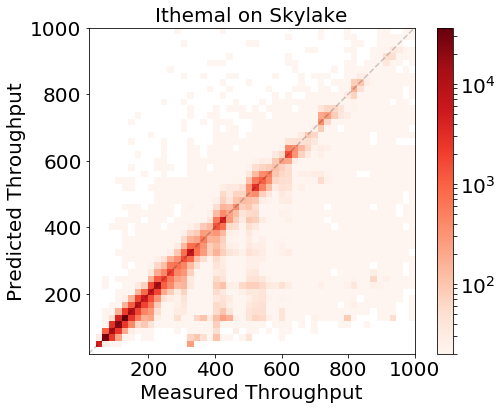}
	}
        \subfloat[][llvm-mca]{
		\includegraphics[width=0.33\textwidth]{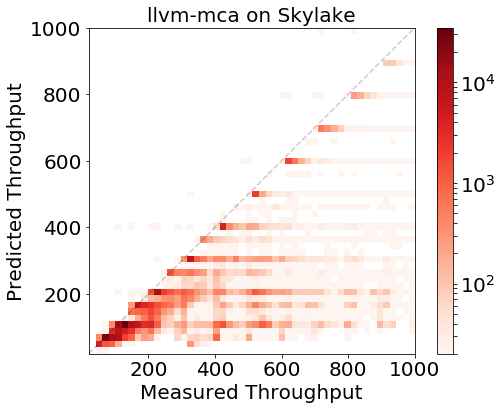}
	}
        \subfloat[][IACA]{
		\includegraphics[width=0.33\textwidth]{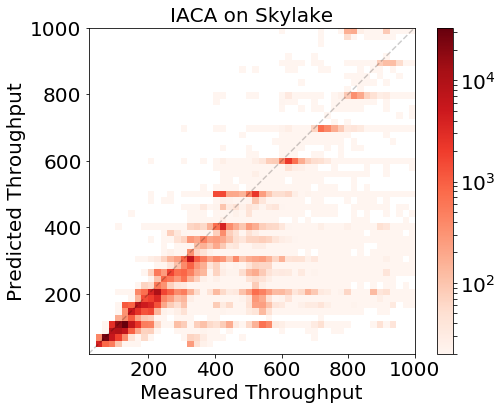}
	}

	\caption{Heatmaps for measured and predicted throughput values under different models for basic blocks with measured throughput values less than 1000 cycles for the Intel Ivy Bridge, Haswell and Skylake microarchitectures}
	\label{fig:heatmaps}
\end{figure*}

Figure~\ref{fig:heatmaps} shows all prediction heatmaps for \tool{}, llvm-mca and IACA under the Intel Ivy Bridge, Haswell and Skylake microarchitectures.
Note that the latest IACA version does not support Ivy Bridge and hence its prediction heatmap is not available.

\section{Prediction Errors for Throughput Ranges}
\label{app:error}

\begin{figure*}
	\centering
	\subfloat[][Ivy Bridge - error curve]{
		\includegraphics[width=0.33\textwidth]{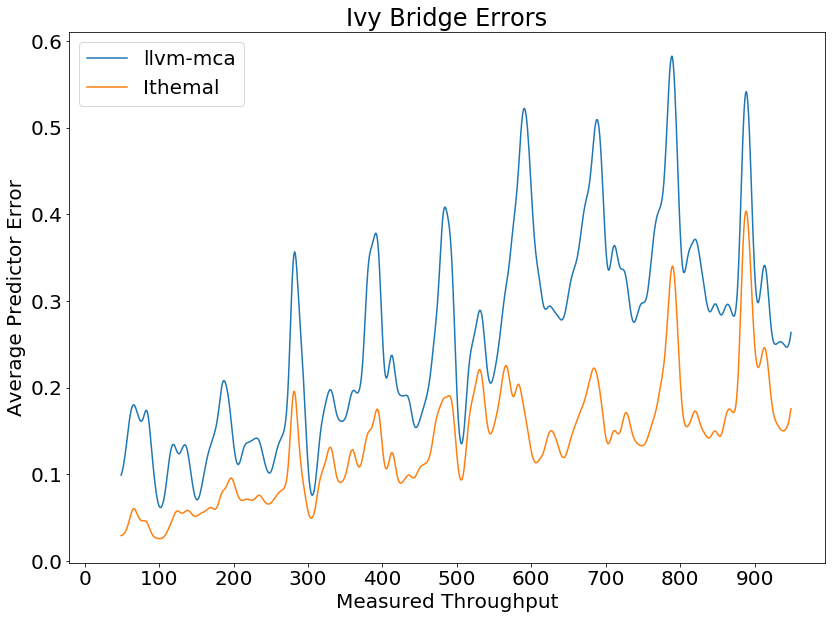}
	}
        \subfloat[][Ivy Bridge - throughput distribution]{
		\includegraphics[width=0.33\textwidth]{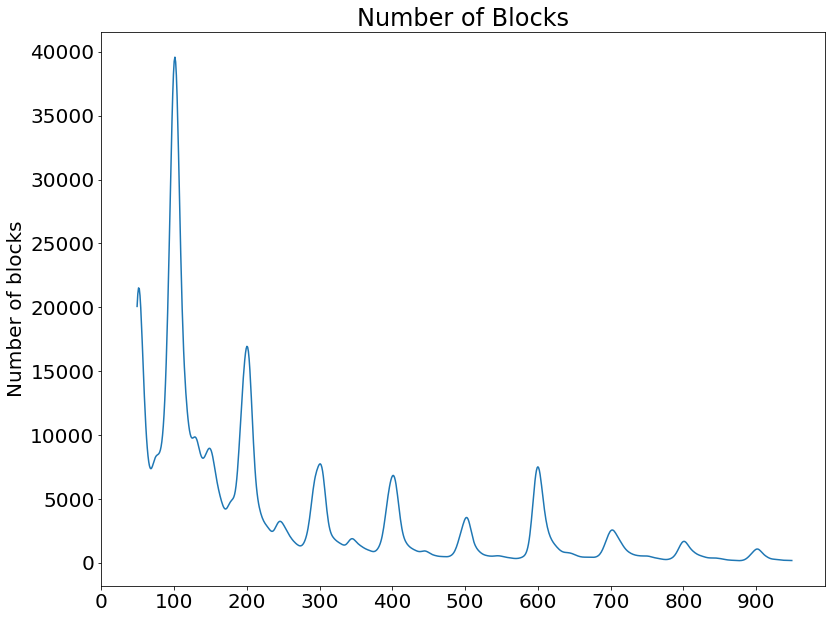}
	}
        \subfloat[][Ivy Bridge - best predictor percentage]{
		\includegraphics[width=0.33\textwidth]{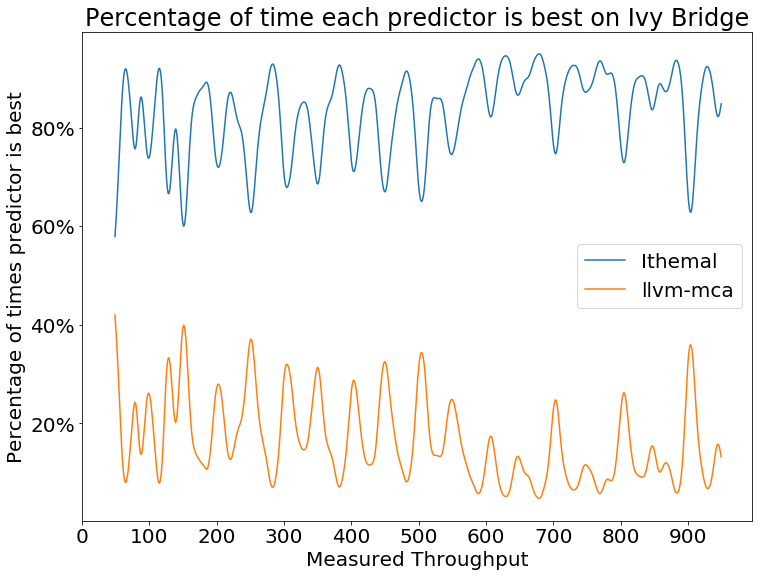}
	}\\
	\subfloat[][Haswell - error curve]{
		\includegraphics[width=0.33\textwidth]{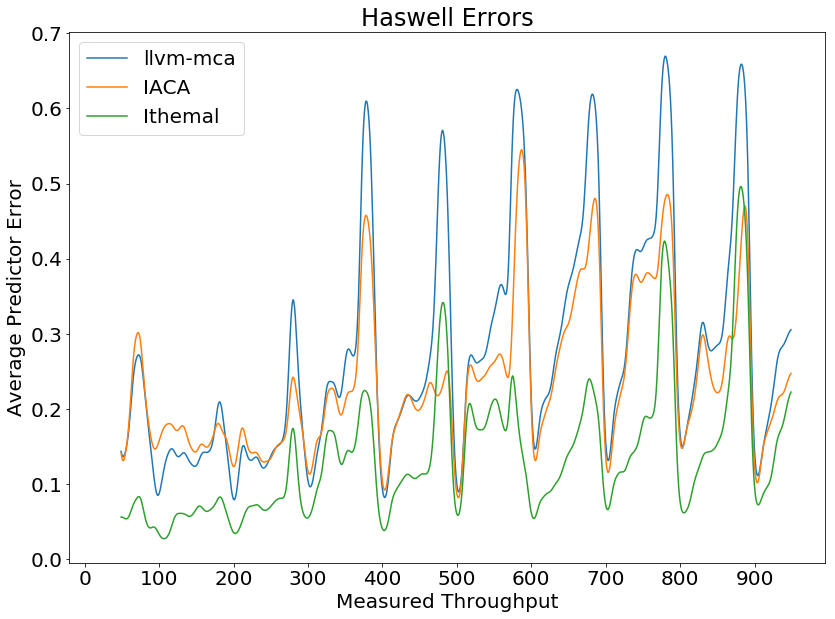}
	}
        \subfloat[][Haswell - throughput distribution]{
		\includegraphics[width=0.33\textwidth]{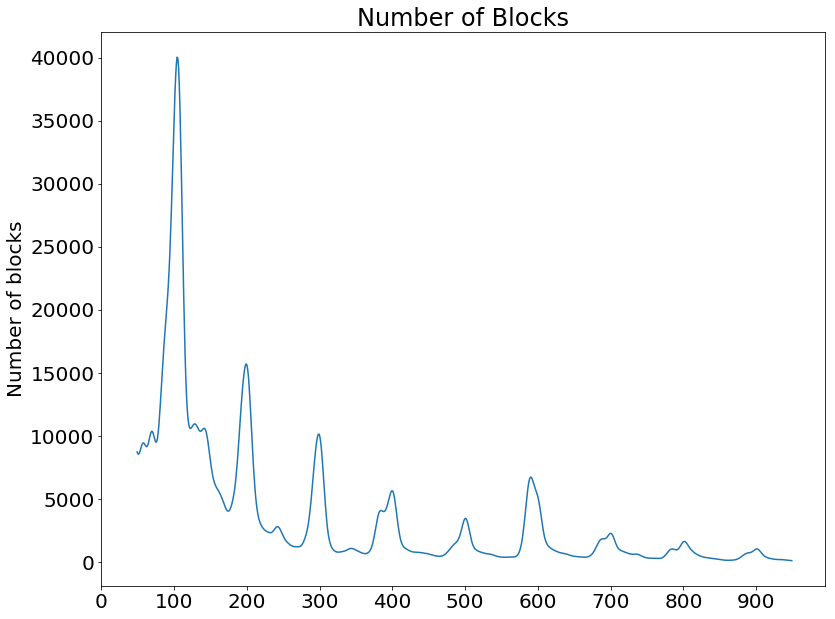}
	}
        \subfloat[][Haswell - best predictor percentage]{
		\includegraphics[width=0.33\textwidth]{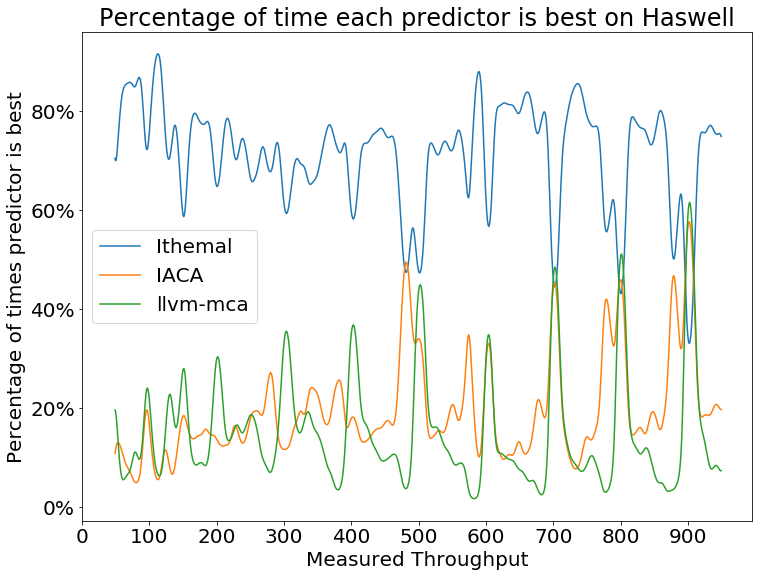}
	}\\
	\subfloat[][Skylake - error curve]{
		\includegraphics[width=0.33\textwidth]{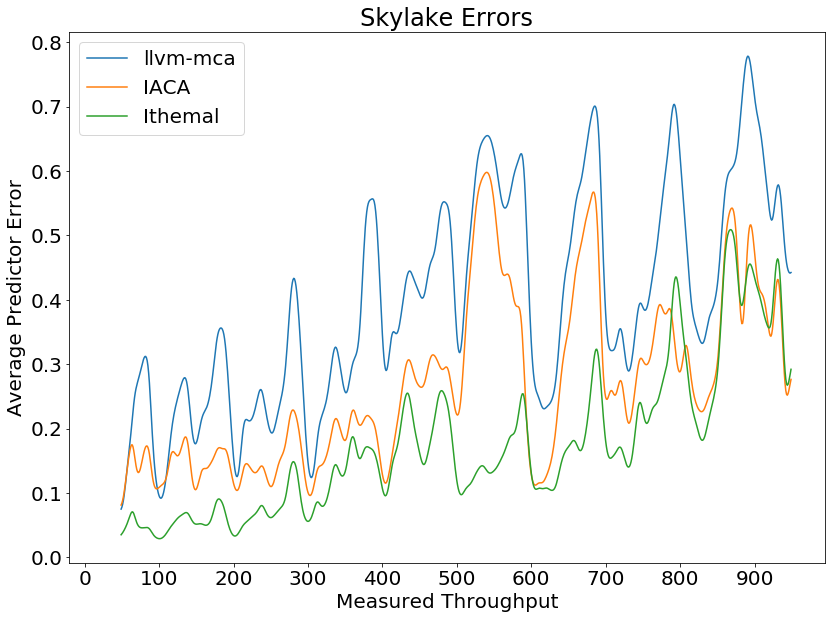}
	}
        \subfloat[][Skylake - throughput distribution]{
		\includegraphics[width=0.33\textwidth]{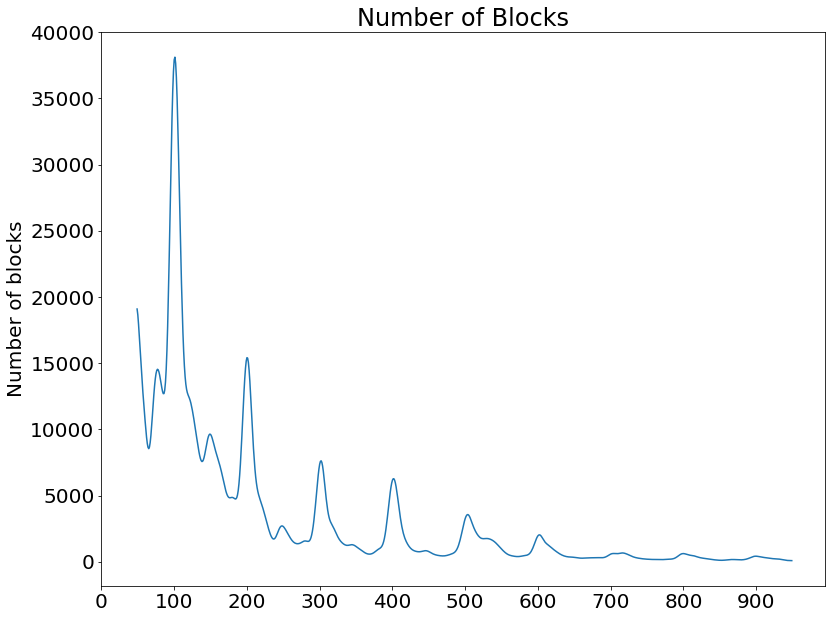}
	}
        \subfloat[][Skylake - best predictor percentage]{
		\includegraphics[width=0.33\textwidth]{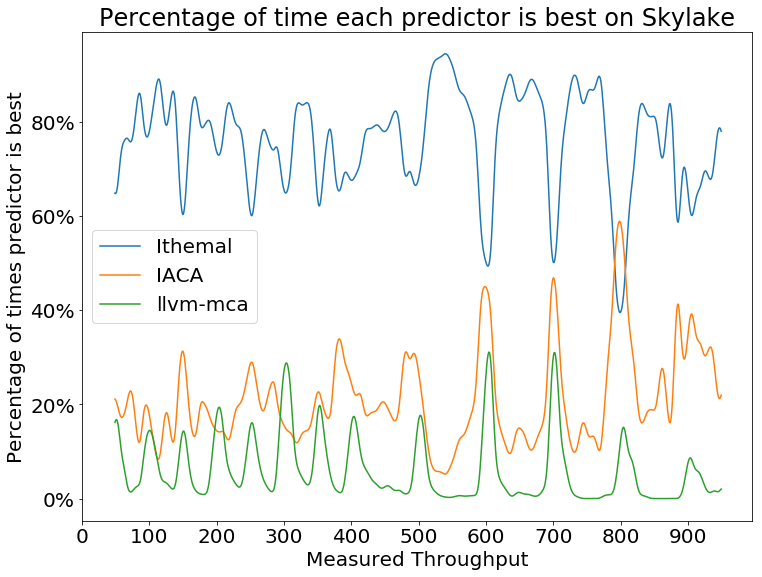}
	}

	\caption{Average error curves and throughput distributions for different models for basic blocks with measured throughput values less than 1000 cycles under different microarchitectures for the test set}
	\label{fig:errorall}
      \end{figure*}

Figure~\ref{fig:errorall} shows how the average error changes between various throughput ranges for each prediction method under different microarchitectures for basic blocks with throughput values under 1000 cycles.
Throughput values are broken up in to bins of length and width 20 cycles on each axis.
It also shows the throughput distribution of the basic blocks, and the average error across different measured throughput ranges.
\tool{} consistently predicts throughput values with lower average errors compared to llvm-mca and IACA.
Overall, \tool{} is more robust in its prediction across all throughput ranges compared to llvm-mca and IACA which show higher fluctuations.

\section{Token RNN Architecture}
\label{app:token-rnn-architecture}

\begin{figure*}[h!]
    \centering
    \includegraphics[width=\textwidth]{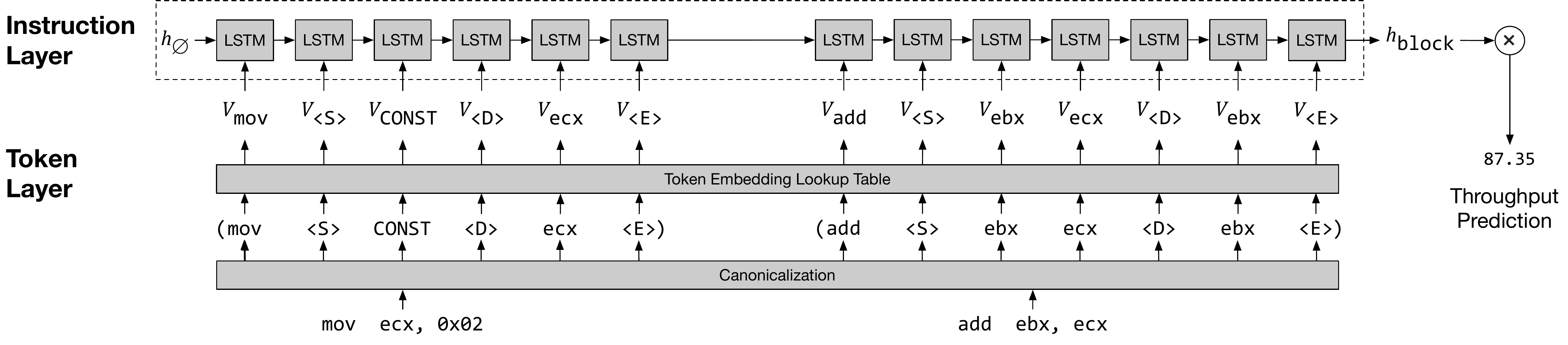}
    \vspace{-20pt}
    \caption{Token RNN Architecture}
    \label{fig:token-rnn-architecture}
\end{figure*}

The full architecture for the Token RNN as presented in Section~\ref{sec:exploration} is shown in Figure~\ref{fig:token-rnn-architecture}.

